\documentclass[runningheads]{llncs}
\usepackage{graphicx}

\title{
    Prediction of a T-cell/MHC-I-based immune profile for colorectal liver metastases from CT images using ensemble learning}

\author{
    Ralph Saber$^{1,2}$,
    David Henault$^{2,3}$,   
    Rolando Rebolledo$^{2,3}$,
    Simon Turcotte$^{2,3}$,
    Samuel Kadoury$^{1,2}$}

\authorrunning{R. Saber et al.}
\titlerunning{Ensemble TabNet predicting an immune profile from CT images}
 \institute{
    $^{1}$ Polytechnique Montr\'eal, Montreal, Quebec, Canada \\
    $^{2}$ Centre de recherche du CHUM, Montreal, Quebec, Canada
    \\
    $^{3}$ Universit\'e de Montr\'eal, Montreal, Quebec, Canada
    }

\begin{document}


\maketitle

\begin{abstract}
Colorectal cancer liver metastases (CLM) are the most common type of distant metastases originating from the abdomen and are characterized by a high recurrence rate after curative resection. It has been previously reported that CLM presenting a low cluster of differentiation 3 (CD3) positive T-cell infiltration density concurrent with a high major histocompatibility complex class I (MHC-I) expression were associated with poor clinical outcomes. In this study, we attempt to noninvasively predict whether a CLM exhibit the CD3$^{Low}$MHC$^{High}$ immunological profile using preoperative CT images. To this end, we propose an ensemble network combining multiple Attentive Interpretable Tabular learning (TabNet) models, trained using CT-derived radiomic features. A total of 160 CLM were included in this study and randomly divided between a training set (n=130) and a hold-out test set (n=30). The proposed model yielded good prediction performance on the test set with an accuracy of 70.0\% [95\% confidence interval 53.6\%-86.4\%] and an area under the curve of 69.4\% [52.9\%-85.9\%]. It also outperformed other off-the-shelf machine learning models. We finally demonstrated that the predicted immune profile was associated with a shorter disease-specific survival ($p$ = .023) and time-to-recurrence ($p$ = .020), showing the value of assessing the immune response. 
\end{abstract}

\begin{keywords}
Colorectal cancer liver metastases, noninvasive prognostic biomarker, immune profiling, attentive interpretable tabular learning, ensemble learning.
\end{keywords}

\section{Introduction}
\label{sec:intro}
Colorectal cancer (CRC) ranks second in terms of cancer-related mortality worldwide \cite{Rebersek2021-lu}. 
Colorectal liver metastases (CLM) are the most common distant metastases of colorectal origin. In 30 to 40\% of patients with metastatic CRC, CLM develop as the exclusive metastatic site \cite{Valderrama-Trevino2017-ba}. Amongst CLM patients who meet the resectability criteria, 20 to 50\% survive for more than five years after curative resection \cite{10.1158/1078-0432.CCR-21-0163}. Nevertheless, postsurgical recurrence rates are close to 70\% \cite{Buisman2020-zq}. Several efforts have been made to understand the underlying biological processes involved in tumor genesis and identify prognostic biomarkers.

Evidence shows that cancer outcomes do not solely depend on the tumoral cells, but also on the microenvironment surrounding the tumor \cite{Van_den_Eynde2018-wj}. Particularly, tumor infiltrating lymphocytes (TIL) have been associated with patient outcomes and metastatic progression \cite{Baldin2020-uo}, and emerged as a novel cancer biomarker. Studies showed a cluster of differentiation 3 (CD3) positive TIL could stratify CRC patients’ survival \cite{doi:10.1126/science.1129139}. Moreover, findings suggest that adding the major histocompatibility complex class I (MHC-I) enhances the prognostic value of CD3+ TIL in CLM \cite{Turcotte2014-rv}. A specific group of patients with CLM characterized by a low CD3+ T-cell infiltration density, concordant with a high MHC-I expression (CD3$^{Low}$MHC$^{High}$) was shown to have shorter disease-specific survival (DSS) and time-to-recurrence (TTR) and poorer response to chemotherapy \cite{ascodavid}.

Several machine learning methods have been proposed for disease stratification \cite{Yu2021-we}, as well as for the prediction of response to therapy \cite{Cha2017-iz}. Recent radiomic studies have shown that imaging features could be leveraged to decipher not only macroscopic patterns, but also hidden biological processes within the tumor microenvironment \cite{Lambin2012-fk}. In this context, several studies have tested the association between radiomic features and the TIL density. Tang et al. \cite{Tang2018DevelopmentOA} separated non-small cell lung cancer (NSCLC) patients into four clusters using CT derived radiomic features and distinguished a class of patients characterized by good outcomes and concurrent low PD-L1 expression and high CD3+ TIL density. Yoon et al. \cite{Yoon2020-ug} proposed to predict stratified densities of type 2 helper T-cells in NSCLC patients by training a linear discriminant analysis model on radiomic features extracted from CT images and reached an area under the curve (AUC) of 68.4\% on the test set. Nevertheless, applications on lung cancer are vastly predominant. The potential of radiomic features extracted from CT images for the prediction of immunological phenotypes of CLM remains largely uninvestigated.

Ensemble learning is a technique that consists in aggregating the outputs of several predictors, hence creating a single composite model. Ensemble learning addresses limitations in small-sized datasets, as they tend to overfit on the training data. Ensembling multiple models allows to mitigate this issue by combining several learned models, leading to higher generalization capability than a single predictor. Furthermore, given that the output of an ensemble model depends on different predictors, it is less likely to get trapped in local minima during training. Moreover, having multiple models working in tandem allows to expand the search space and reduces the risk that the solution falls outside the model’s space. Finally, some ensemble learning algorithms are particularly useful when dealing with high dimensional feature sets or class imbalance \cite{ensemblelearning}.

In this work, we evaluated ensemble feature learning to predict if CLM exhibit the CD3$^{Low}$MHC$^{High}$ immune phenotype from diagnostic CT. To this end, we proposed to train an ensemble of the Attentive Interpretable Tabular Learning (TabNet) model. In a post-hoc analysis, we assessed whether the predicted immunological phenotype could stratify patient outcomes, namely the TTR and the DSS.

\section{Materials and Methods}
\label{sec:methods}


\begin{figure*}[htb]
    \centering
    \begin{minipage}[b]{0.69\linewidth}
      \centering
      \centerline{\includegraphics[width=\linewidth]{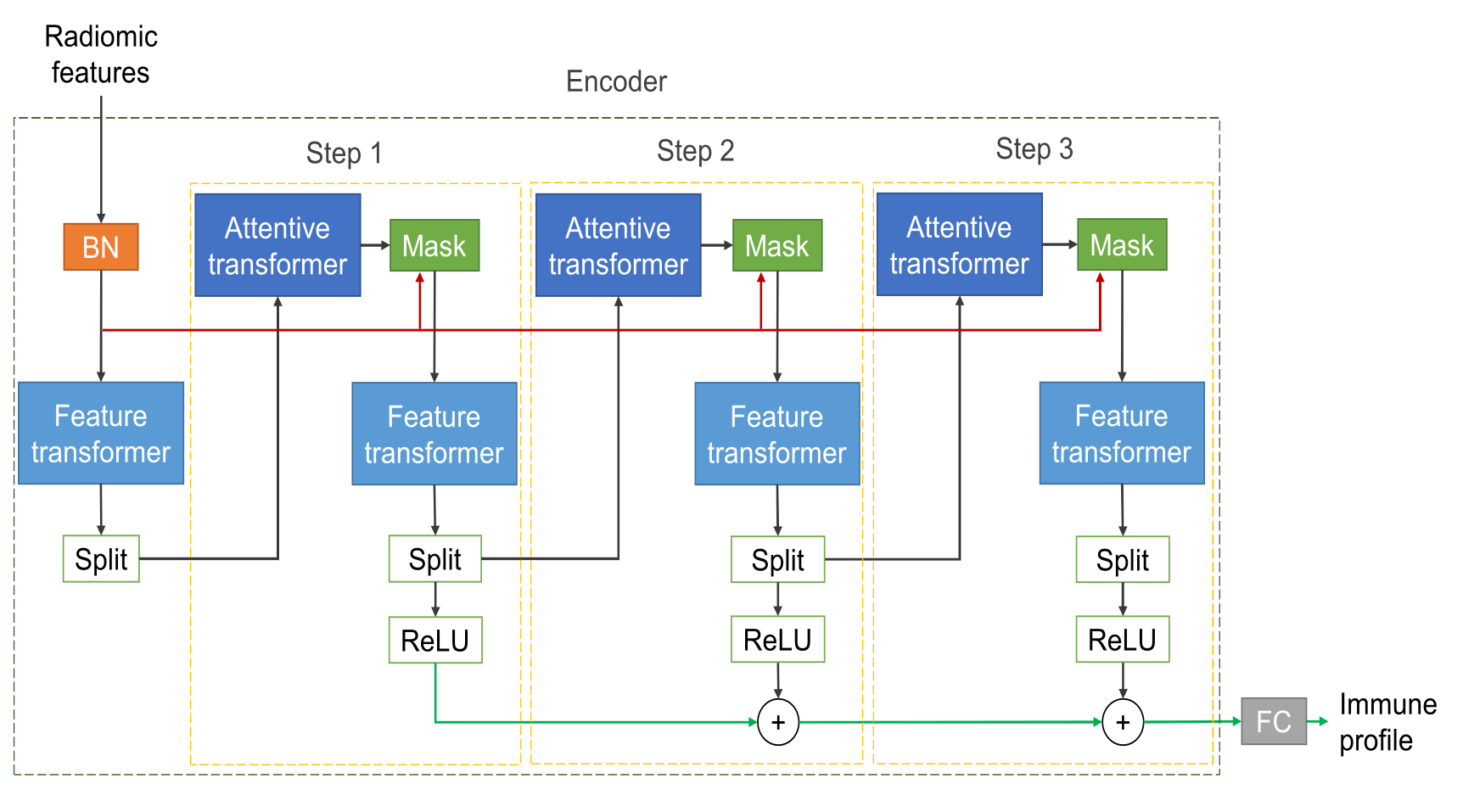}}
      \centerline{(a)}\medskip
    \end{minipage}
    \begin{minipage}[b]{0.29\linewidth}
      \centering
      \centerline{\includegraphics[width=\linewidth]{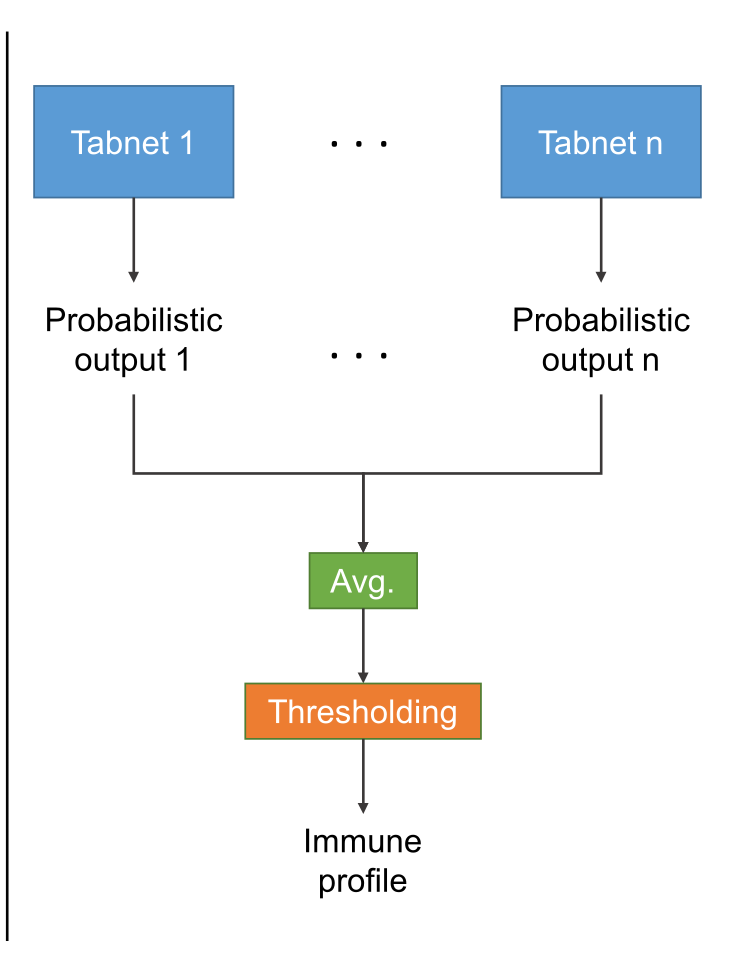}}
      \centerline{(b)}\medskip
    \end{minipage}
    \caption{
        (a) Attentive Interpretable Tabular Learning (TabNet) model architecture. The attentive transformer outputs a feature selection mask which aim at discarding irrelevant features at a given decision step. The feature transformer processes the selected features through shared and decision step-specific layers. The model’s output is obtained by adding a fully connected layer on top of the encoder's output. (b) Proposed Multi-TabNet ensemble. The multi-TabNet ensemble aggregates separately trained TabNet models through voting. Avg, Averaging; BN, Batch Normalization; FC, Fully Connected layer; ReLU, Rectified Linear Unit.}
    \label{fig:model}
\end{figure*}

\begin{figure*}[htb]
    \centering
    \begin{minipage}[b]{0.3\linewidth}
      \centering
      \centerline{\includegraphics[width=\linewidth]{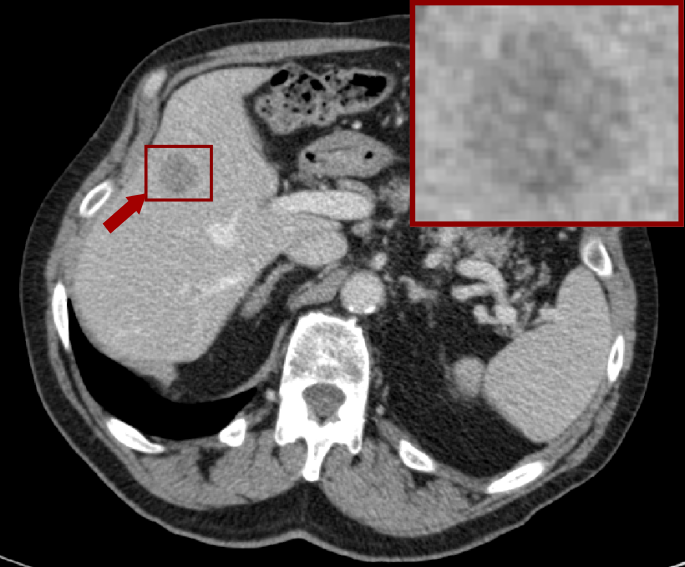}}
      \centerline{(a)}\medskip
    \end{minipage}
    \begin{minipage}[b]{0.3\linewidth}
      \centering
      \centerline{\includegraphics[width=\linewidth]{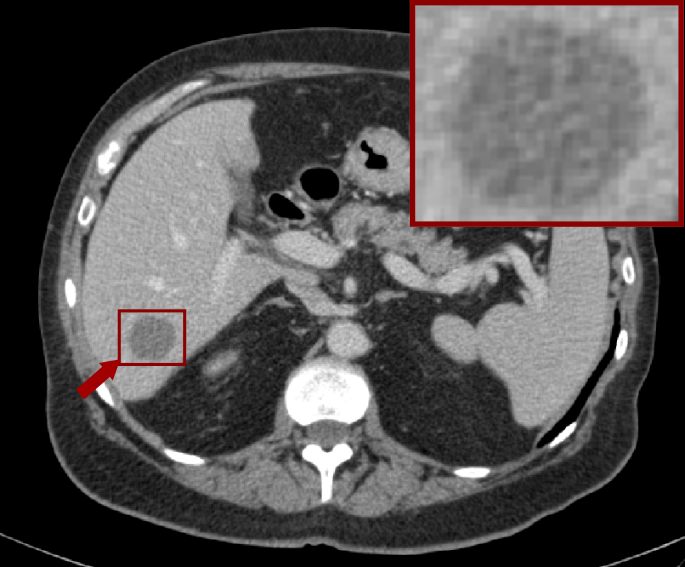}}
      \centerline{(b)}\medskip
    \end{minipage}
    \begin{minipage}[b]{0.3\linewidth}
      \centering
      \centerline{\includegraphics[width=\linewidth]{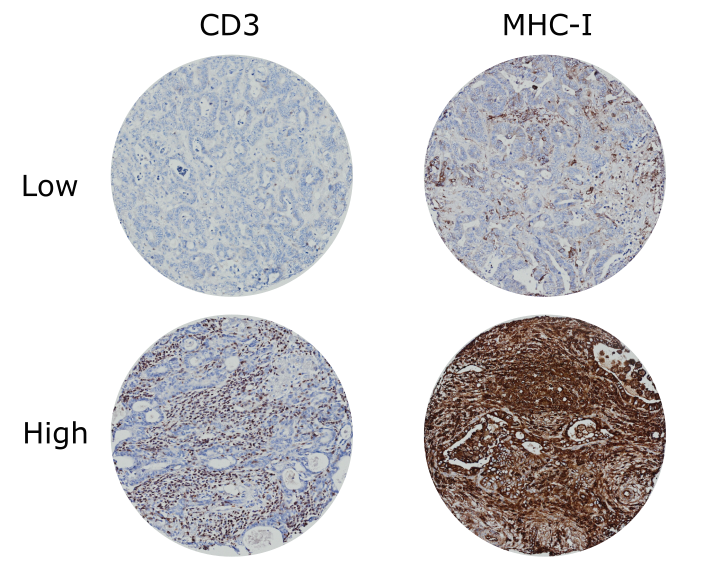}}
      \centerline{(c)}\medskip
    \end{minipage}
    \caption{
        Representative CT images with (a) CD3$^{Low}$MHC-I$^{High}$ CLM and (b) CD3$^{Low}$MHC-I$^{Low}$ CLM. (c) Representative TMA cores with CD3 and MHC-I staining.}
    \label{fig:repfigs}
\end{figure*}

\subsection{Patient cohort}
\label{sec:cohort}

In this study, 160 pathologically confirmed CLM from 122 consenting patients who underwent curative resection at the Centre Hospitalier de l’Universit\'e de Montr\'eal, were used from our prospectively maintained dataset. The cohort had a mean age of 63.4 (35 – 84) and was predominantly constituted of male patients (63.9\%). The associated clinical outcomes, namely the DSS and the TTR were also provided. 

\subsection{Data preparation}
\label{sec:data}

\subsubsection{Imaging data}
\label{sssec:imaging}

Presurgical CT images of CLM patients were included. The images had a transverse volume size of 512 × 512, a mean axial resolution of 0.72 mm$^2$ (range = [0.56, 0.98] mm$^2$) and a mean slice thickness of 2.26 mm (range = [0.80, 5.0] mm). An isotropic spatial resampling was applied in order to have a uniform voxel spacing across samples (1×1×1mm$^3$). A joint liver/hepatic automatic lesion segmentation algorithm \cite{Vorontsov2018LiverLS} was utilized to segment the CLM. The segmentation model consists of two convolutional networks: the former segments the liver and the latter the lesions within. Segmentation masks were then validated by an experienced radiologist.

\subsubsection{Immunological data}
\label{sssec:immuno}

In this study, we used previously produced CLM immunological data \cite{ascodavid}. For every CLM, tissue microarrays (TMA) were prepared as described in \cite{Messaoudi2022-kk} using six 0.6 mm diameter intratumoral core biopsies so as to take into account tumor heterogeneity and ensure sufficient immunological representation of the entire metastatic lesion. The TMA was immunohistochemically stained for CD3 (F7.2.38, Dako, Carpinteria, CA) and MHC-I (HC10, mouse monoclonal pan-anti‐human HLA class I, provided by H. Ploegh, Whitehead Institute, Cambridge, MA, USA), digitized at ×20 and automated counting of CD3 and MHC-I was performed, yielding the number of CD3+ cells per mm$^2$ and the ratio of MHC-I positive surface area over total assessable area, respectively. Subsequently, CD3 and MHC-I values were stratified into low and high groups, by setting a cutoff of 932.1 cells/mm$^2$ and 45.0\% for CD3 and MHC-I, respectively. The cutoff was chosen by the X-tile software \cite{xtile} with the technique adopted for colorectal cancer immune scoring using the minimal $p$ value approach with outcome data \cite{Messaoudi2022-kk}. Lesions were finally divided into 2 classes according to whether they concurrently exhibit CD3$^{Low}$ and MHC-I$^{High}$ profiles (64 lesions) or not (96 lesions).

\subsection{Radiomic feature extraction}
\label{sec:radiomic}

From each segmented CLM, 107 radiomic features were extracted using the PyRadiomics v3.0.1 toolbox \cite{pyradiomics}. The rationale behind using radiomic features is twofold: (1) radiomic pipelines tend to yield better results on small-sized datasets than end-to-end models which are inherently data-hungry and prone to overfitting when trained on limited data; (2) radiomic features have the distinctive advantage of being interpretable, which is particularly useful in the medical field where transparency is mandatory for the deployment in routine clinical practice. Features consisted of 18 first-order statistics, 14 shape, and 75 textural features. All features were standardized in order to obtain a feature set with a mean of zero and a standard deviation of one.

\subsection{Prediction framework for T-cell/MHC-I profile}
\label{sec:modeldesc}

In this study, we propose a multi-TabNet ensemble model to predict whether a given CLM exhibits the CD3$^{Low}$MHC$^{High}$ immunological profile (Fig. \ref{fig:model}).

\subsubsection{Attentive Interpretable Tabular Learning}
\label{sssec:tabnet}

The Attentive Interpretable Tabular Learning (TabNet) model is a multi-stage deep learning model introduced by Google Cloud AI \cite{tabnet} and applies a sequential instance-wise attention mechanism allowing it to inherently select the most salient set of radiomic features at different decision steps within its architecture (Fig. \ref{fig:model}). TabNet was shown to outperform the Extreme Gradient Boosting (XGBoost) classifier, previously considered as the state-of-the-art model for tabular data.

The model incorporates two key components: an attentive transformer and a feature transformer. The former aims at generating feature selection masks, which are used to discard irrelevant features at each step. For this, the attentive transformer processes features from a previous step through a single fully connected layer. It then generates sparse probabilities using a sparsemax activation while taking into account the extent to which each feature has contributed to the previous step. The feature transformer processes the ensuing filtered features using two subsequent blocks of layers: shared and decision step-dependent layers. Each block consists of two successive sets of fully connected layers, batch normalization layers (BN) and gated linear units (GLU).

The radiomic features are first sent to a BN layer and processed using a feature transformer block. Then at each step, a feature selection mask is generated by the attentive transformer and multiplied with the full feature set in order to obtain the optimal set of features in the given step. The masking operation is decision step-specific, meaning that a different set of features is selected at each step. The filtered features are subsequently processed by the feature transformer and divided into two parts: the first part is sent to the subsequent decision step while the second one is used as the output of the current decision step.

The encoder output is a linear combination of the decision step output and Tabnet’s output is computed by applying a fully connected layer to the encoder’s output.

\subsubsection{Multi-TabNet ensemble model}
\label{sssec:multitabnet}

In this work, we propose to compute the probabilistic outputs of multiple TabNet models trained separately for the prediction of the CD3/MHC-I immunological profile and then aggregate the outputs using soft voting \cite{softvoting} (Fig. \ref{fig:model}).

In hard voting, the class with the highest number of votes is returned by the ensemble model whereas in soft voting, the class probabilities $P$ of each TabNet model $T_j$, trained on the input \textbf{x}, are averaged and the class $i$ with the highest ensuing probability is picked by the model (Eq. 1):

\begin{equation}
    \hat{y} = argmax_{i}(\frac{1}{n}\sum_{j=1}^{n}P(T_j(x)=i)))
\end{equation}

where $\hat{y}$ is the predicted class and $n$ is the number of models in the ensemble.

\section{Results and discussion}
\label{sec:results}

\begin{figure}[ht]
    \begin{minipage}{\linewidth}
      \centering
      \centerline{\includegraphics[width=\linewidth]{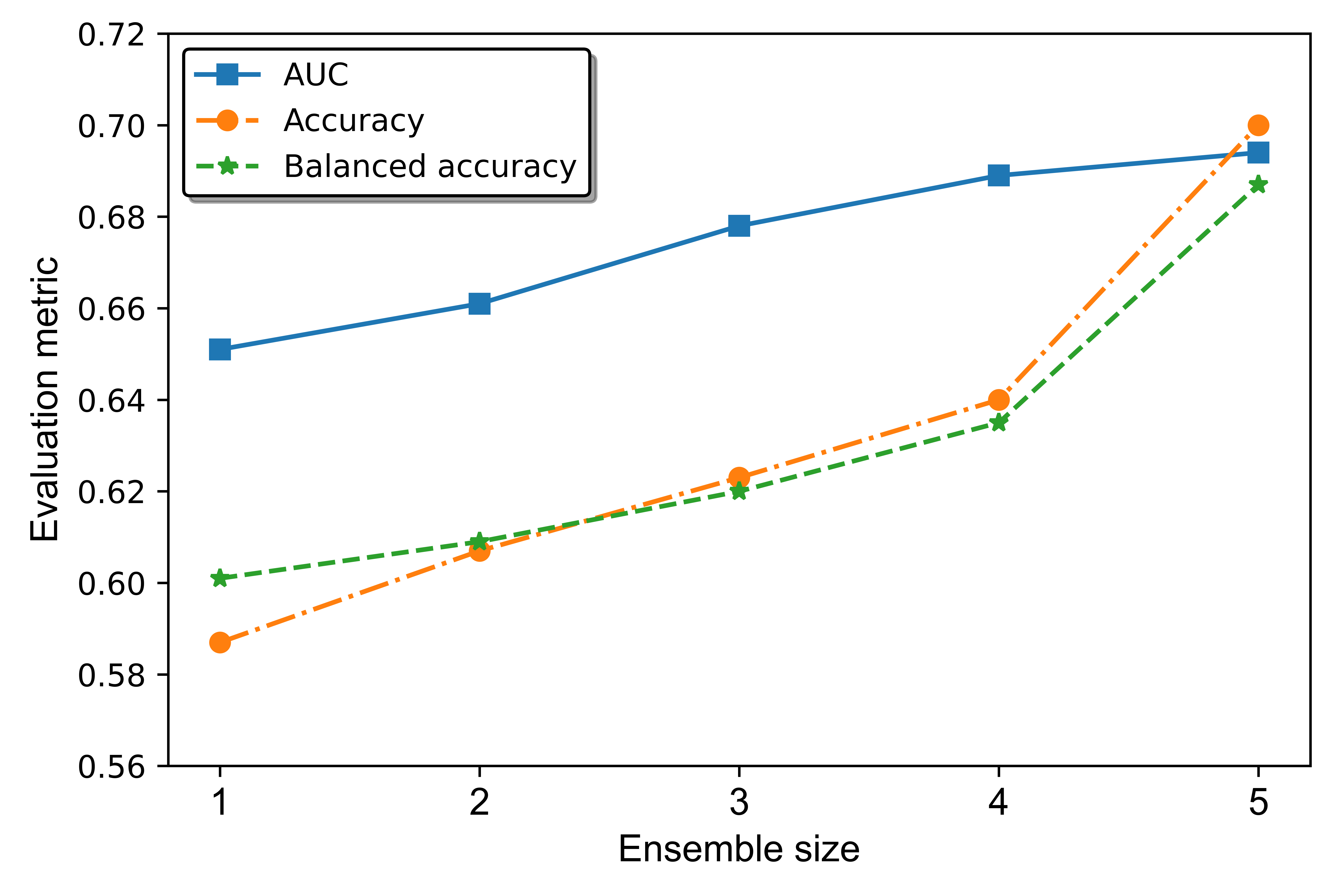}}
    \end{minipage}
    \caption{Evolution of Multi-TabNet ensemble performance in terms of AUC, accuracy and balanced accuracy, by ensemble size (n). For n$<$5, the average performance across all possible combinations of n-sized ensembles is reported.
        }
    \label{fig:ensembleeffect}
\end{figure}

\begin{figure*}[htb]
    \centering
    \begin{minipage}[b]{0.45\linewidth}
      \centering
      \centerline{\includegraphics[width=\linewidth]{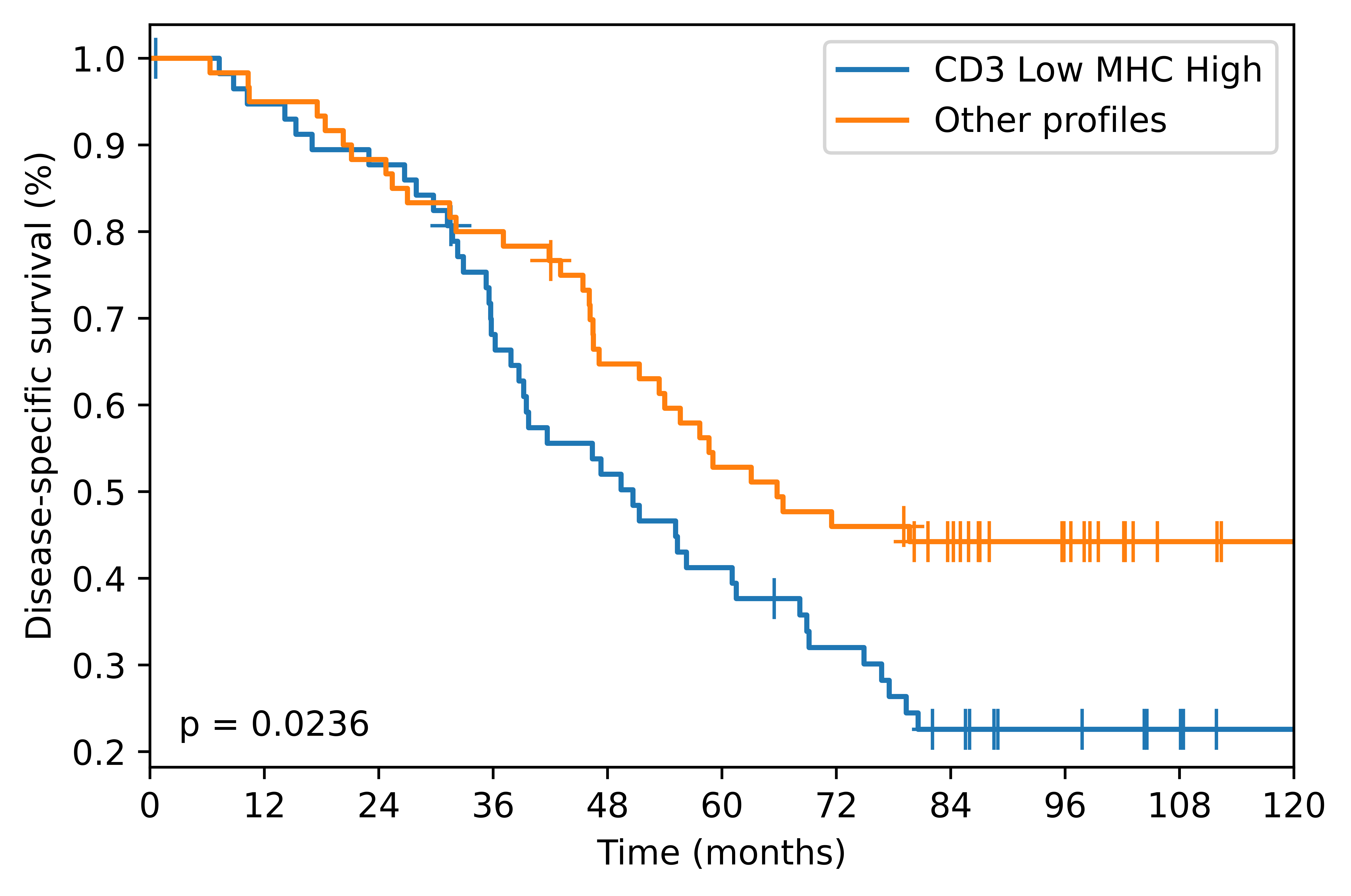}}
     \centerline{(a)}\medskip
    \end{minipage}
    \begin{minipage}[b]{0.45\linewidth}
      \centering
      \centerline{\includegraphics[width=\linewidth]{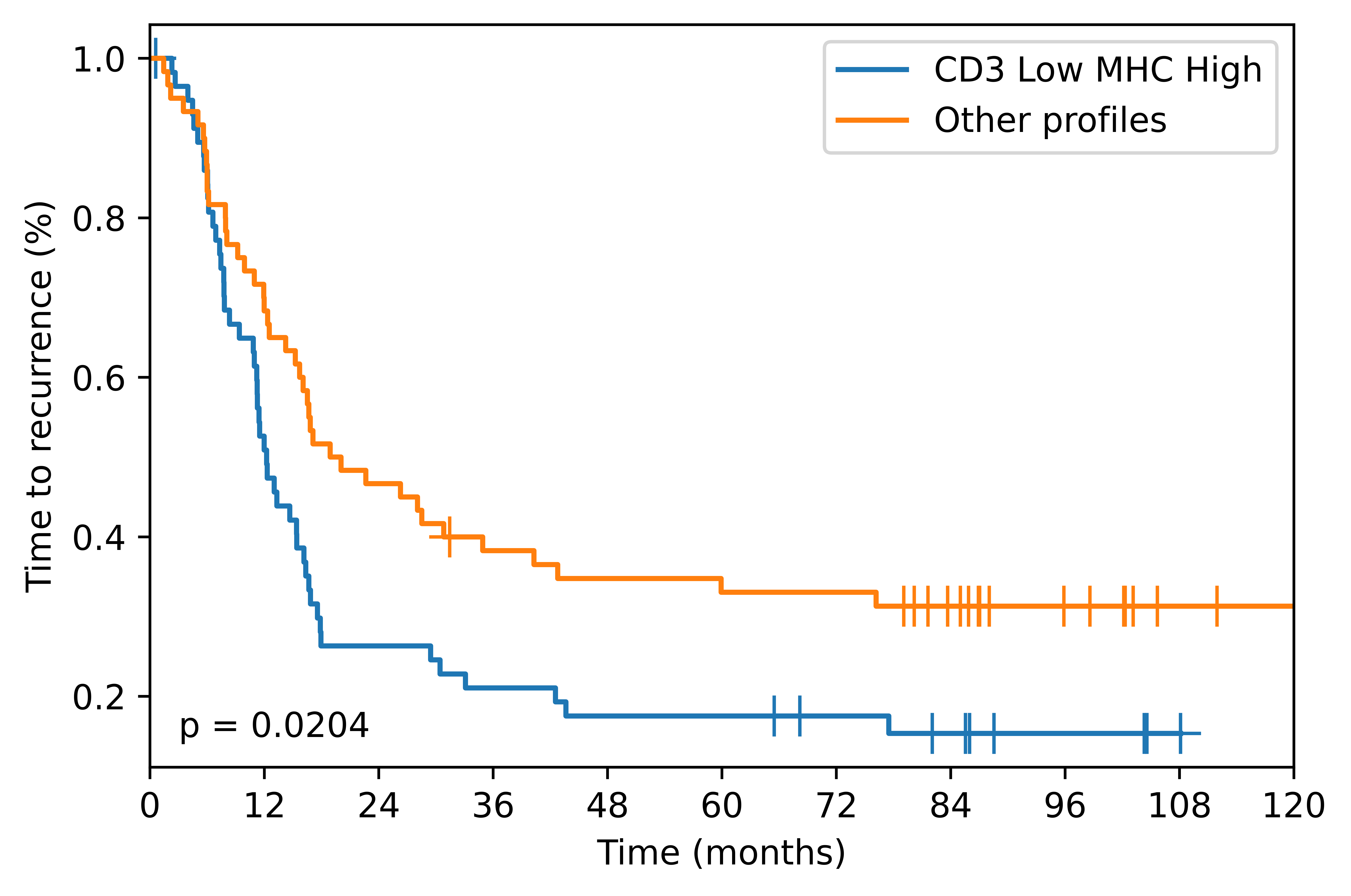}}
     \centerline{(b)}\medskip
    \end{minipage}
    \caption{
        Prognostic value of the predicted immune profile: (a) DSS and (b) TTR following the initial CLM resection.}
    \label{fig:kmcurvers}
\end{figure*}

\subsection{Evaluation methodology}
\label{sec:experiments}

The dataset was randomly divided between a training subset (130 lesions) and a hold-out test subset (30 lesions). To avoid any information crossing between the training and the hold-out test sets, all lesions of a given patient were always included in the same subset.

For benchmarking purposes, we also trained a logistic regression model (LR), a linear support vector machine (SVM) and an XGBoost classifier. For all trained models, a 5-fold cross-validation scheme was applied on the training set for hyperparameter tuning. To handle class imbalance, the Synthetic Minority Over-sampling Technique (SMOTE) \cite{smote} was used on the training set prior to training. All models were trained to predict whether CLM exhibit the CD3$^{Low}$MHC$^{High}$ profile or not (Fig. \ref{fig:repfigs}).

The TabNet ensemble model was trained using an effective batch size of 64 and the Adam optimizer with a binary cross-entropy loss function for 100 epochs. Training was performed on an NVIDIA GeForce GTX TITAN Xp 12GB. The performance metrics adopted were the AUC, sensitivity, specificity, accuracy and balanced accuracy which takes into account the number of instances in each class. 

In order to study the ensembling effect on TabNet models performance, we tested multi-TabNet ensembles with sizes going from one to five.

\subsection{Prediction of the CD3/MHC-I-based immune profile}
\label{sec:predresults}

\begin{table}[ht]
    \centering
    \caption{
        Performance on the hold-out test set of the different models on the prediction of the CD3/MHC-I-based immune profile. Top results are shown in bold. 95\% confidence intervals are represented between brackets.}
    \scalebox{1}{
    \begin{tabular}{c @{\hspace{1\tabcolsep}} c @{\hspace{1\tabcolsep}} c @{\hspace{1\tabcolsep}} c @{\hspace{1\tabcolsep}} c @{\hspace{1\tabcolsep}} c}
         \hline
         Model & {AUC(\%)} & {Acc(\%)} & {Sens(\%)} & {Spec(\%)} \\
         \hline\hline
         LR & {48.3 [30.4-66.2]} & 53.3 & 54.5 & 52.6 \\
         \hline
         SVM & {64.1 [47.0-81.3]} & 60.0 & \textbf{63.6} & 57.9 \\
         \hline
         XGBoost & {68.9 [52.3-85.5]} & 63.3 & 54.5 & 68.4 \\
         \hline
         \textbf{Multi-Tabnet} & \textbf{69.4 [52.9-85.9]} & \textbf{70.0} & \textbf{63.6} & \textbf{73.7} \\
         \hline
    \end{tabular}
    }
    \label{tab:benchmark}
\end{table}

Table \ref{tab:benchmark} depicts the results of the different models trained for the prediction of the immunological status: CD3$^{Low}$MHC$^{High}$ versus other profiles. The results show that the proposed multi-Tabnet ensemble combining 5 TabNet models outperforms the other machine learning models with an AUC of 69.4\% [95\% confidence interval 52.9\%-85.9\%], an accuracy of 70.0\% [53.6\%-86.4\%] and a balanced accuracy of 68.7\% [52.1\%-85.3\%]. It is to be noted that the proposed model exhibits a satisfying detection of both classes as mirrored by the sensitivity and specificity values.

As shown in Fig. \ref{fig:ensembleeffect}, when the number of aggregated TabNet models in the proposed architecture increases from 1 to 5, the accuracy, the balanced accuracy and the AUC increase from 58.7\% to 70.0\%, 60.1\% to 68.7\% and 65.1\% to 69.4\%, respectively. This demonstrates the benefit of ensembling on the prediction performance.

\subsection{Prognostic value of the predicted immune profile}
\label{sec:prognosisresults}

With the goal of assessing whether the predicted immunological profile is associated with patient outcomes, we used the Kaplan-Meier technique to draw the survival curves (DSS and TTR) up to ten years post-surgery of patients having at least one CD3$^{Low}$MHC$^{High}$ lesion versus the rest of the cohort. Patient stratification into the two groups was performed by setting a cut-off value on the probabilistic output of the ensemble model, chosen using the X-tile software \cite{xtile}. Survival curves were compared using the log-rank test.

The Kaplan-Meier curves of Fig. \ref{fig:kmcurvers} show that patients with at least one CLM with a CD3$^{Low}$MHC$^{High}$ profile had statistically significantly shorter DSS ($p$ = .023) and TTR ($p$ = .020) than the rest of the patients. This reveals that the CT-based immune profiling of the CLM holds prognostic value and could serve as a preoperative biomarker.

\section{Conclusion}
\label{sec:conclusion}

In this study, we demonstrated that a transformer-based ensemble network trained with extracted radiomic features from preoperative CT images, hold relevant biological information that could be associated with CLM immunological profiling. The proposed multi-TabNet ensemble network was capable of noninvasively identifying lesions exhibiting a low CD3+ TIL density and a high MHC-I expression. Moreover, experimental results showed that the proposed model yielded good results on the hold-out test set and outperformed the other machine learning models trained on the same task. Finally, we showed that the predicted profile was associated with patients’ outcomes and could potentially be used as a noninvasive prognostic indicator of CLM. Prospectively, one could focus on training the model on a larger sample size and performing an external validation to evaluate its generalizability across different centers and patient ethnicities. Furthermore, with more data available, one could attempt to make use of the CD3/MHC-I data to perform a four-class classification and assess whether each of the four classes translates into a different clinical phenotype.

\section{Compliance with ethical standards}
\label{sec:ethics}

This study was performed in line with the principles of the Declaration of Helsinki. Approval was granted by the Institutional Review Board.

\section{Acknowledgments}
\label{sec:acknowledgments}

This work was supported by the Canada Research Chairs and by the National Science and Engineering Research Council of Canada and the Université de Montréal Roger Des Groseillers Research Chair in Hepatopancreatobiliary Surgical Oncology. ST and SK are scientists of the Centre de recherche du Centre hospitalier de l’Université de Montréal (CRCHUM) supported by the Fonds de recherche du Québec - Santé (FRQ-S). ST was supported by the FRQ-S Young Clinician Scientist Seed Grant (No. 32633), the FRQS Clinician Scientist Junior-1 and 2 Salary Award (No. 30861, No. 298832), and the Institut du Cancer de Montréal establishment award. DH was supported by the FRQ-S phase 1 award for medical resident engaged in clinician-scientist training. The authors have no financial or non-financial interests to disclose.

\bibliographystyle{IEEEbib}
\bibliography{bibliography2}

\begin{thebibliography}{10}

\bibitem{Rebersek2021-lu}
Martina Rebersek,
\newblock ``Gut microbiome and its role in colorectal cancer,''
\newblock {\em BMC Cancer}, vol. 21, no. 1, pp. 1325, Dec. 2021.

\bibitem{Valderrama-Trevino2017-ba}
Alan~I Valderrama-Trevi{\~n}o, Baltazar Barrera-Mera, Jes{\'u}s~C
  Ceballos-Villalva, et~al.,
\newblock ``Hepatic metastasis from colorectal cancer,''
\newblock {\em Euroasian J Hepatogastroenterol}, vol. 7, no. 2, pp. 166--175,
  Sept. 2017.

\bibitem{10.1158/1078-0432.CCR-21-0163}
Preeti Kanikarla~Marie, Cara Haymaker, Edwin~Roger Parra, et~al.,
\newblock ``{Pilot Clinical Trial of Perioperative Durvalumab and Tremelimumab
  in the Treatment of Resectable Colorectal Cancer Liver Metastases},''
\newblock {\em Clinical Cancer Research}, vol. 27, no. 11, pp. 3039--3049, 06
  2021.

\bibitem{Buisman2020-zq}
Florian~E Buisman, Boris Galjart, Eric~P van~der Stok, et~al.,
\newblock ``Recurrence patterns after resection of colorectal liver metastasis
  are modified by perioperative systemic chemotherapy,''
\newblock {\em World J Surg}, vol. 44, no. 3, pp. 876--886, Mar. 2020.

\bibitem{Van_den_Eynde2018-wj}
Marc Van~den Eynde, Bernhard Mlecnik, Gabriela Bindea, et~al.,
\newblock ``The link between the multiverse of immune microenvironments in
  metastases and the survival of colorectal cancer patients,''
\newblock {\em Cancer Cell}, vol. 34, no. 6, pp. 1012--1026.e3, Dec. 2018.

\bibitem{Baldin2020-uo}
Pamela Baldin, Marc Van~den Eynde, Bernhard Mlecnik, et~al.,
\newblock ``Prognostic assessment of resected colorectal liver metastases
  integrating pathological features, {RAS} mutation and immunoscore,''
\newblock {\em J Pathol Clin Res}, vol. 7, no. 1, pp. 27--41, Sept. 2020.

\bibitem{doi:10.1126/science.1129139}
Jérôme Galon, Anne Costes, Fatima Sanchez-Cabo, et~al.,
\newblock ``Type, density, and location of immune cells within human colorectal
  tumors predict clinical outcome,''
\newblock {\em Science}, vol. 313, no. 5795, pp. 1960--1964, 2006.

\bibitem{Turcotte2014-rv}
Simon Turcotte, Steven~C Katz, Jinru Shia, et~al.,
\newblock ``Tumor {MHC} class {I} expression improves the prognostic value of
  t-cell density in resected colorectal liver metastases,''
\newblock {\em Cancer Immunol Res}, vol. 2, no. 6, pp. 530--537, Feb. 2014.

\bibitem{ascodavid}
David Henault, David Stephen, Pierre-Antoine St-Hilaire, et~al.,
\newblock ``Prognostic immune scoring of colorectal cancer liver metastasis
  with mhc class-i expression combined to t cell quantification.,''
\newblock {\em Journal of Clinical Oncology}, vol. 36, no. 15\_suppl, pp.
  3586--3586, 2018.

\bibitem{Yu2021-we}
Hang Yu, Laurence~T Yang, Qingchen Zhang, et~al.,
\newblock ``Convolutional neural networks for medical image analysis:
  State-of-the-art, comparisons, improvement and perspectives,''
\newblock {\em Neurocomputing}, vol. 444, pp. 92--110, July 2021.

\bibitem{Cha2017-iz}
Kenny~H Cha, Lubomir Hadjiiski, Heang-Ping Chan, et~al.,
\newblock ``Bladder cancer treatment response assessment in {CT} using
  radiomics with {Deep-Learning},''
\newblock {\em Scientific Reports}, vol. 7, no. 1, pp. 8738, Aug. 2017.

\bibitem{Lambin2012-fk}
Philippe Lambin, Emmanuel Rios-Velazquez, Ralph Leijenaar, et~al.,
\newblock ``Radiomics: extracting more information from medical images using
  advanced feature analysis,''
\newblock {\em Eur J Cancer}, vol. 48, no. 4, pp. 441--446, Jan. 2012.

\bibitem{Tang2018DevelopmentOA}
Chad Tang, Brian~Paul Hobbs, Ahmed~M Amer, et~al.,
\newblock ``Development of an immune-pathology informed radiomics model for
  non-small cell lung cancer,''
\newblock {\em Scientific Reports}, vol. 8, 2018.

\bibitem{Yoon2020-ug}
Hyun~Jung Yoon, Jun Kang, Hyunjin Park, et~al.,
\newblock ``Deciphering the tumor microenvironment through radiomics in
  non-small cell lung cancer: Correlation with immune profiles,''
\newblock {\em PLoS One}, vol. 15, no. 4, pp. e0231227, Apr. 2020.

\bibitem{ensemblelearning}
Omer Sagi and Lior Rokach,
\newblock ``Ensemble learning: A survey,''
\newblock {\em WIREs Data Mining and Knowledge Discovery}, vol. 8, no. 4, pp.
  e1249, 2018.

\bibitem{Vorontsov2018LiverLS}
Eugene Vorontsov, Gabriel Chartrand, An~Tang, et~al.,
\newblock ``Liver lesion segmentation informed by joint liver segmentation,''
\newblock {\em 2018 IEEE 15th International Symposium on Biomedical Imaging
  (ISBI 2018)}, pp. 1332--1335, 2018.

\bibitem{Messaoudi2022-kk}
Nouredin Messaoudi, David Henault, David Stephen, et~al.,
\newblock ``Prognostic implications of adaptive immune features in
  {MMR-proficient} colorectal liver metastases classified by histopathological
  growth patterns,''
\newblock {\em British Journal of Cancer}, vol. 126, no. 9, pp. 1329--1338, May
  2022.

\bibitem{xtile}
Robert~L Camp, Marisa Dolled-Filhart, and David~L Rimm,
\newblock ``X-tile: a new bio-informatics tool for biomarker assessment and
  outcome-based cut-point optimization,''
\newblock {\em Clin Cancer Res}, vol. 10, no. 21, pp. 7252--7259, Nov. 2004.

\bibitem{pyradiomics}
Joost J.~M. van Griethuysen, Andriy~Y Fedorov, Chintan Parmar, et~al.,
\newblock ``Computational radiomics system to decode the radiographic
  phenotype.,''
\newblock {\em Cancer research}, vol. 77 21, pp. e104--e107, 2017.

\bibitem{tabnet}
Sercan~{\"O}. Arik and Tomas Pfister,
\newblock ``Tabnet: Attentive interpretable tabular learning,''
\newblock {\em ArXiv}, vol. abs/1908.07442, 2021.

\bibitem{softvoting}
Saloni Kumari, Deepika Kumar, and Mamta Mittal,
\newblock ``An ensemble approach for classification and prediction of diabetes
  mellitus using soft voting classifier,''
\newblock {\em International Journal of Cognitive Computing in Engineering},
  vol. 2, pp. 40--46, June 2021.

\bibitem{smote}
Nitesh~V. Chawla, Kevin~W. Bowyer, Lawrence~O. Hall, et~al.,
\newblock ``{SMOTE}: Synthetic minority over-sampling technique,''
\newblock {\em Journal of Artificial Intelligence Research}, vol. 16, pp.
  321--357, jun 2002.

\end{thebibliography}

\AtEndDocument{\vfill%
© 2023 IEEE.  Personal use of this material is permitted.  Permission from IEEE must be obtained for all other uses, in any current or future media, including reprinting/republishing this material for advertising or promotional purposes, creating new collective works, for resale or redistribution to servers or lists, or reuse of any copyrighted component of this work in other works.}

\end{document}